\documentclass{article}

\usepackage{amssymb,amsmath}
\usepackage{fullpage}
\usepackage{graphicx}
\usepackage{subfigure}
\usepackage{appendix}
\usepackage{amsmath}	
\usepackage{slashed}
\usepackage[margin=1.0in]{geometry}
\usepackage{setspace}
\usepackage{color}
\usepackage{fancyhdr}
\usepackage{collcell}
\usepackage{datatool}
\usepackage{environ}
\usepackage{latexsym}
\usepackage{amssymb}
\usepackage{epsfig,amsmath,graphics}
\usepackage{epstopdf}
\usepackage{verbatim}
\usepackage{wasysym}
\usepackage{feynmp-auto}
\usepackage{authblk}
\usepackage{xcolor}
\usepackage{enumitem}
\usepackage[utf8]{inputenc}
\usepackage{slashed}
\usepackage{cite}
\newcommand{\tm}[1]{\textcolor{blue}{ #1 }}

\begin{document}
\title{What can solve the Strong CP problem?}







 \date{\today}

\author[1]{David E.~Kaplan}
\author[2]{Tom Melia}
\author[1]{Surjeet Rajendran}
\affil[1]{\small Department of Physics \& Astronomy, The Johns Hopkins University, Baltimore, MD  21218, USA}
\affil[2]{\small Kavli IPMU (WPI), UTIAS, The University of Tokyo, Kashiwa, Chiba 277-8583, Japan}
\maketitle

\begin{abstract}

Three possible strategies  have been advocated to solve the strong CP problem. The first is the axion, a dynamical mechanism that relaxes any initial value of the CP violating angle $\bar{\theta}$ to zero. The second is the imposition of new symmetries that are believed to set $\bar{\theta}$ to zero in the UV. The third is the acceptance of the fine tuning of parameters.  We argue that the latter two solutions do not solve the strong CP problem. The $\theta$ term of QCD is not a parameter - it does not exist in the Hamiltonian. Rather, it is a property of the quantum state that our universe finds itself in, arising from the fact that there are CP violating states of a CP preserving Hamiltonian. It is not eliminated by imposing parity as a symmetry since the underlying theory is already parity symmetric and that does not preclude the existence of CP violating states. Moreover, since the value of $\theta$ realized in our universe is a consequence of measurement, it is inherently random and cannot be fine tuned by choice of parameters. Rather any fine tuning would require a tuning between parameters in the theory and the random outcome of measurement. Our results considerably strengthen the case for the existence of the axion and axion dark matter.  The confusion around $\theta$ arises from the fact that unlike classical mechanics, the Hamiltonian and Lagrangian are not equivalent in quantum mechanics. The Hamiltonian defines the differential time evolution, whereas the Lagrangian is a solution to this evolution.  Consequently, initial conditions could in principle appear in the Lagrangian but not in the Hamiltonian. This results in aspects of the initial condition such as $\theta$ misleadingly appearing in the Lagrangian as parameters.   We comment on the similarity between the $\theta$ vacua and the violations of the constraint equations of classical gauge theories in quantum mechanics. 

\end{abstract}

\tableofcontents

\section{Introduction}

Experimental searches for the electric dipole moment of nucleons have constrained the electric dipole moment to be smaller than $d_{n, p} \lessapprox 10^{-26}$ e-cm. It is experimentally interesting to search for this term since an electric dipole moment can exist for fundamental particles only when CP symmetry is broken. The experimental limit is puzzling since there are at least two separate sources of CP violation that contribute to this dipole moment. The first is a phase that might exist in the quark mass matrix. The second arises from the $\theta$ term of QCD. The sum of these two terms gives rise to an effective CP violating phase $\bar{\theta}$. The experimental limit on the nucleon electric dipole moment implies $\bar{\theta} \lessapprox 10^{-9}$. Given that these two terms should naively be $\mathcal{O}\left(1\right)$ numbers, why should they cancel so precisely? The Strong CP problem is aimed at understanding the resolution to this fine tuning problem. 

The most popular solution to this problem is the axion\cite{Peccei:1977hh, Peccei:1977ur, Weinberg:1977ma, Wilczek:1977pj}. The axion is a dynamical solution - irrespective of the value of $\bar{\theta}$, the axion adjusts its field value so as to cancel the net CP violation in the theory, including the electric dipole moment of nucleons. But, additional solutions to the strong CP problem have also been proposed. These solutions leverage the fact that $\bar{\theta}$ is radiatively stable - that is, if it was set to zero at some high scale, then standard model contributions to it are extremely small and not in conflict with observation. These solutions attempt to solve the strong CP problem by imposing parity (or CP) as a symmetry of the theory\cite{Barr:1984qx,Craig:2020bnv,Dunsky:2019api,Feruglio:2023uof,Hall:2024xbd,Nelson:1983zb}. It is believed that the imposition of these  symmetries  independently forbids the existence of CP violating phases in the quark mass matrix and sets $\theta = 0$. Since both these terms are independently set to zero, $\bar{\theta}$ starts from zero and the radiative stability of this term then ensures that there is no conflict with observation. Another possibility that is sometimes advocated is to simply accept the observed smallness of $\bar{\theta}$ as an initial fine tuning, even if one cannot associate lofty anthropic philosophies with the smallness of $\bar{\theta}$.

The main point of this paper is to argue that the strong CP problem can only be solved through dynamical methods such as the axion and not by either invoking parity (or CP) as a symmetry of the theory or by fine tuning. This is fundamentally tied to the fact that the $\theta$ term of QCD is {\bf not} a parameter of the theory that can be set to zero by imposition of symmetries on the Hamiltonian. Rather it reflects a choice of vacuum state or eigenstate of the QCD vacuum. For a generic initial state, as the universe evolves from early times, the value of $\theta$ is randomly picked in our universe -- {\it i.e.}, it is a consequence of measurement. Thus $\theta$ is set probabilistically. Cancellation of its effects require the choice of parameters in the Hamiltonian to exactly cancel this random outcome of measurement.  Moreover, the Hamiltonian of QCD preserves CP - the imposition of parity does not add any additional content to the theory. But, a CP preserving Hamiltonian can have states (or vacua) that break CP. It is the existence of such states that are the origin of $\theta$.

Why is there confusion about these points in the literature? This is due to the fact that when the stong CP problem is often discussed, it is characterized as a term in the Lagrangrian where it looks like a parameter, giving rise to the belief that it can be set to zero by imposing symmetries on (or fine tuning) the Lagrangian. In classical physics, the Lagrangian and Hamiltonian are dual descriptions of the same physical system. Parameters of the Lagrangian are also parameters of the Hamiltonian. In quantum mechanics this is not the case - the Hamiltonian is an operator which defines the time evolution of the quantum mechanical system. The Lagrangian of course appears in defining the path integral -- but, the path integral is a solution to the time evolution operator. Solutions to differential equations require initial conditions. Thus, the Lagrangian can have terms that are reflective of the choice of initial states to perform the time evolution. These terms are not determined by the symmetries of the Hamiltonian and thus cannot be set to zero by imposing restrictions on the Hamiltonian. 

The purpose of this paper is two-fold. First, we wish to provide a pedagogic description of the above physics with illustrative examples. We note that Gia Dvali has articulated the above position many times over the past decades ({\it e.g.}, \cite{Dvali:2005zk}). These issues are also described in other publications such as \cite{Jackiw:1983nv, CALLAN1976334}. But, these points are not widely appreciated. Here we offer a concise argument that illustrates these points.  Second, we highlight the connection between the $\theta$ term of QCD and recent work \cite{Burns:2022fzs, Kaplan:2023fbl, Kaplan:2023wyw, DelGrosso:2024gnj} on the absence of constraints in gauge theories. As we will show, in both cases, the proposed physics ($\theta$ and constraint violations) does not exist at the level of the Hamiltonian. Rather, when specific solutions to the time evolution are constructed, the choice of initial conditions enters the Lagrangian as apparent parameter choices. 

The rest of this paper is organized as follows. In Section \ref{sec:hydrogen}, we discuss these issues in the simple quantum mechanical case of the hydrogen atom. This discussion is tied to the problem of choosing states of a theory - but there is no connection to this system and the strong CP problem. We then discuss the case of Bloch waves in a periodic crystal where $\theta$  arises as a choice of energy eigenstate. In this section, we also show how $\theta$ enters the Lagrangian seemingly as a parameter even though it does not appear in the Hamiltonian. The $\theta$ problem is also sometimes described in terms of the quantum mechanics of a rigid pendulum swinging under gravity. Similarities between the periodic crystal and the pendulum are discussed in appendix \ref{appendix}. We apply these lessons to QCD in Section \ref{sec:QCD}. The similarity between the $\theta$ vacua and the absence of constraints in gauge theories is discussed in Section \ref{sec:gauge}. These discussions are of significant import in the search of axion dark matter - these are discussed in our conclusions in Section \ref{sec:conclusions}. 

\section{Hydrogen Atom}
\label{sec:hydrogen}

Consider the spherically symmetric Hamiltonian that describes the hydrogen atom: 

\begin{equation}
H_{H} = -\frac{1}{2 \mu_e^2} \nabla^2 - \frac{\alpha}{r}
\label{Eqn:MasslessH}
\end{equation}

Here $\mu_e$ is the reduced mass of the atom and $\alpha$ is the fine structure constant. As is well known, even though the Hamiltonian \eqref{Eqn:MasslessH} is spherically symmetric, there are a number of energy eigenstates of \eqref{Eqn:MasslessH} that are not spherically symmetric such as the 2p, 3d, and other excited states. In our universe, when we find a random hydrogen atom, we expect it to be in the ground state, $1$s. We are not surprised by this even though we know that in our universe, the protons and electrons were initially at very high temperatures and thus when atoms form during recombination, we expect that the electrons will land in arbitrary bound states of the hydrogen atom. However, the excited states of hydrogen have very short lifetimes and thus they rapidly decay to the ground state. 

Now suppose we lived in a different universe where the photon had a mass $m_{\gamma}$ with the mass being between $\alpha^2 \mu_e < m_{\gamma} < \alpha \mu_e $. The Hamiltonian for this system is 

\begin{equation}
H_{m} = -\frac{1}{2 \mu_e^2} \nabla^2 - \frac{q^2}{r} e^{-m_{\gamma}r}
\label{Eqn:MassiveH}
\end{equation}
When $\alpha^2 \mu_e < m_{\gamma} < \alpha \mu_e $, the range of the photon is longer than the Bohr radius $(\alpha \mu_e)^{-1}$. This implies that there are some number of bound states of the electron in this system. For example, we expect some number of low lying eigenstates of the form 1s, 2s, 2p, {\it etc.}, to exist, as long as their Bohr radii are shorter than the range of the electric field. But, when $m_{\gamma} > \alpha^2 \mu_e$ these excited states cannot decay to the ground state by emitting photons because the binding energy $\alpha^2 \mu_e$ is smaller than the energy needed to emit the massive photon. 

In this world, if we knew that the protons and electrons had initially been at very high temperatures, when that universe undergoes recombination, we would expect it to populate a number of these energy eigenstates of $H_{m}$. But, since these low lying levels cannot decay, we would expect to find a population of states in the excited states such as 2s, 2p {\it etc}. If observations are made and we only find atoms in the 1s state, we would be confused - why were the higher energy states that are stable not populated? 

In this world, there would then be a ``1s'' problem with particle physicists devising methods to solve this problem. What kinds of solutions are possible? Notice we cannot solve this problem by imposing spherical symmetry on the Hamiltonian and thus believing that this imposition would get rid of states like 2p and 3d. This is because $H_m$ is already spherically symmetric and this spherically symmetric Hamiltonian nevertheless has non-spherically symmetric states. Could we imagine some kind of ``fine tuning'' apparatus that exists at every point in space to bring the electron to the ground state? What kind of fine tuning mechanism would we need? When the protons and electrons come from high temperature and the bound states are formed, the formation of a specific bound state is due to measurement - that is, the observer gets entangled with various outcomes of the high energy recombination process and probabilistically ends up in some specific state. Thus, at each point  (where the atom exists), we probabilistically obtain one of the many possible stable eigenstates. The fine tuning mechanism would thus have to be dynamic - it would need to look at the specific outcome that was probabilistically obtained and send that specific state to the ground state.

We recognize that the particle physicists in this world need to solve this problem through dynamical mechanisms - they can for example invoke the existence of a light or massless hidden photon that would allow the metastable states to decay or include additional operators in the theory such as collisions which would allow for inelastic relaxation. 

\section{Periodic Crystal}
\label{sec:pendulum}

 Now consider the example of a particle in a periodic potential, as is the case in a crystal.  The position of the particle is denoted by $x$ and the lattice spacing is $l$. The potential is of the form $V\left(x\right) = -\kappa \cos\left(x/l\right)$.  The quantum mechanical Hamiltonian that describes this system is: 

\begin{equation}
H_{P} = -\frac{1}{2 m} \frac{\partial^2}{\partial x^2} +  V\left(x\right)
\end{equation}

\begin{figure*}[t!]
\centering 

\includegraphics[width=.5\linewidth]{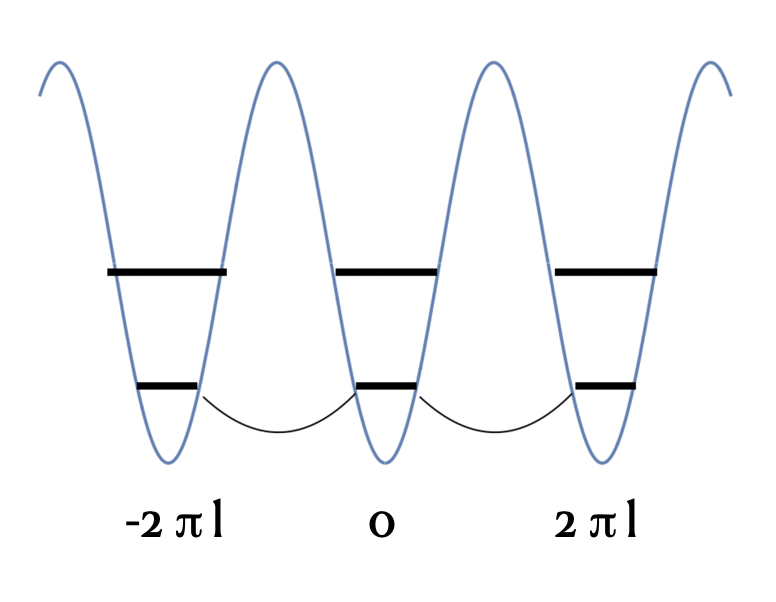}

\caption{The potential $V\left(x\right) = - \kappa \cos\left(x/l\right)$. In the absence of quantum tunneling, each well of the potential is independent of the other wells. Each well would have its own set of energy levels that are exactly degenerate with the corresponding energy levels in the other wells. In the presence of tunneling, the levels mix with each other and produce a band structure. There is an analogous structure to this potential in QCD where we can map the positions  $ x = 0, \dots 2 n \pi l$ to gauge field configurations that are 0 and non-zero pure gauge configurations. } \label{Fig:Band}
\end{figure*}

What are the energy eigenstates of this Hamiltonian? Naively, one might expect the energy eigenstates to be a series of states that are localized around the potential well (see Figure \ref{Fig:Band}). Due to the periodicity of the potential, one might expect a large degeneracy of states with each state localized at different values around the potential at $x =  2  \pi n l$ for integer $n$. However, this is not the correct picture -- due to quantum tunneling, these degenerate states all mix together and form a band structure. The lowest lying levels form the lowest band, with the next highest levels forming the next band and, with high enough barriers, a band gap existing between these two bands.

The energy eigenstates of this system are described by Bloch waves. These waves are constructed by observing that the Hamiltonian commutes with the discrete spatial translation operator $T$ that maps states $|x\rangle \rightarrow |x+ 2\pi l\rangle$. This implies eigenstates of $H_P$ are also eigenstates of $T$ - thus the energy eigenstates $|\Psi_{\theta}\rangle$ are of the form $T |\Psi_{\theta}\rangle = e^{-i \theta} |\Psi_{\theta}\rangle$. $\theta$ here is a label that identifies a particular energy eigenstate just like the labels 1s, 2p {\it etc} in the case of the hydrogen atom. $\theta$ being a choice of energy eigenstate is not a parameter in the theory - it does not appear in $H_P$. 

Now imagine that we have a series of such crystals that were formed in the early universe at some high energy. In this universe, would we expect to find all the crystals in the lowest energy eigenstate in the state with $\theta = 0$? No - in general, we expect to populate all the eigenstates. With this expectation, if we actually went and did an experiment with these crystals and found that all the crystals were in fact in $\theta  = 0$, we will be confused - why did the higher energy states not get populated? Once again, notice that this problem cannot be solved by imposing a symmetry like parity, {\it i.e.}, invariance under $x\rightarrow -x$. The Hamiltonian is already parity symmetric - just like the hydrogen atom was spherically symmetric. But even the parity invariant Hamiltonian contains states that are no longer invariant under parity. Similarly, just like the case of the hydrogen atom, if all these crystals started at some high energy, then the specific low energy state that they will find themselves in is the probabilistic outcome of measurement/entanglement. There is thus no fine tuning mechanism that can rigidly set all the $\theta$ terms to zero for all these crystals. The only way to solve this problem would be through some dynamical mechanism where even though the eigenstates with $\theta \neq 0$ are populated, the state somehow relaxes to $\theta = 0$.

Let us now describe the dynamics of this theory using the path integral. The time evolution operator of this theory is $U = \text{exp}\left(-i H_P t\right)$. This operator can be fully described by providing its transition matrix elements on a set of basis elements. Choosing the position basis and following the canonical procedure to derive the path integral for this operator, one obtains the formula: 

\begin{equation}
\langle x_f | U\left(t\right) | x_i \rangle = \int_{x\left(0\right) = x_i}^{x\left(t\right) = x_f} Dx \,  \text{exp}^{\left(i \int_{\tau = 0}^{\tau = t} d\tau \left(\frac{1}{2} m \dot{x}^2  - V\left(x\right)\right)\right)}
\end{equation}

This Lagrangian does not contain any $\theta$ term. But, clearly, given a solution to the Schr\"{o}dinger equation we have to be able to time evolve any initial state. This is resolved by writing an arbitrary initial state $|\Psi\left(0\right)\rangle$ also in the position basis: 
\begin{equation}
|\Psi\left(0\right\rangle) = \int \,dx_i\, \tilde{\Psi}\left(x_i\right) |x_i\rangle
\end{equation}

We then have: 

\begin{equation}
\langle x_f| U\left(t\right) |\Psi\left(0\right)\rangle = \int dx_i \tilde{\Psi}\left(x_i\right) \langle x_f | U |x_i\rangle
\label{Eqn:NoThetaL}
\end{equation}

Thus, we can pick any initial state we want and obtain the time evolution, including states where $\theta \neq 0$. In these computations, the Lagrangian does not depend on $\theta$ - the information about $\theta$ lives in the initial state wave-function $\tilde{\Psi}\left(x_i\right)$, and random initial state wave function should contain a linear combination of eigenstates of $T$.

We can now ask a different question - suppose we were only interested in understanding the behavior of a specific set of eigenstates  that all transform with the same eigenvalue $\theta$ under spatial translations: $T|\Psi_\theta\rangle = e^{-i \theta} |\Psi_\theta\rangle$. How can we understand just the behavior of this specific sub-space? Since $T$ commutes with the $H_P$, time evolution cannot take a quantum state that has an eigenvalue $\theta$ under $T$ to a state that has a different eigenvalue. In other words, eigenstates of T with a specific value of $\theta$ form a super-selection sector. 

To understand the time evolution of this sector, we can now choose a different position basis, namely: $|[x]_\theta\rangle$ which transforms as $T|[x]_\theta\rangle = e^{-i \theta} |[x]_{\theta}\rangle$. These are the subspaces of the Hilbert space that are invariant under $T$ (all states are multiplied by the same phase). The states $ |[x]_\theta\rangle$ are of the form $|[x]_{\theta}\rangle = e^{i \frac{\theta}{2 \pi} \frac{\hat{x}}{l}} |[x]\rangle$.\footnote{T maps $|x\rangle \rightarrow |x + 2\pi l\rangle$. Thus $ T \hat{x} T^{-1} =  \hat{x} - 2 \pi l$. Hence $T |[x]_\theta\rangle = T e^{i \frac{\theta}{2 \pi}\frac{\hat{x}}{l}}T^{-1} T|[x]\rangle = e^{-i\theta} e^{i \frac{\theta}{2 \pi}\frac{\hat{x}}{l}} |[x]\rangle = e^{-i \theta}|[x]_\theta\rangle$.} defined only  
for $x$ between $x = 0$ and $x = 2 \pi l$ where the state $|[x]\rangle$ is periodic and it is obtained as the sum $|[x]\rangle = \sum_{j} |x + 2\pi j l \rangle$ 
over all integers $j$ (ignoring normalization issues).  Let us now compute the matrix element: 

\begin{equation}
\langle [x_f]_{\theta} | U\left(t\right) |[x_i]_{\theta}\rangle = \langle [x_f]_{\theta} | e^{-i H_P \delta t}\dots e^{-i H_P \delta t} |[x_i]_{\theta}\rangle
\label{Eqn:ThetaPath}
\end{equation}
where we break the time evolution operator into a number of small steps of size $\delta t$. This is the standard way to compute the path integral. After this stage, the standard procedure involves inserting resolutions of the identity operator in terms of the chosen basis. Since $|[x]_{\theta}\rangle$ are eigenstates of the unitary operator $T$,\footnote{Note that T is generated by the momentum operator, which is Hermitean. Eigenstates of this operator are also eigenstates of T. }
we have: 

\begin{equation}
1 = \int_{0}^{2\pi l} dx\int d\theta \,  |[x]_{\theta}\rangle \langle [x]_{\theta}|
\end{equation}

Using this resolution of the identity, \eqref{Eqn:ThetaPath} becomes: 

\begin{equation}
\langle [x_f]_{\theta} | U\left(t\right) |[x_i]_{\theta}\rangle = \langle [x_f]_{\theta} | \dots \int dx_j\int d\theta_{j} |[x_j]_{\theta}\rangle\langle [x_j]_{\theta}| e^{-i H_P \delta t}\int dx_{j-1} d\theta_{j-1} |[x_{j-1}]_{\theta}\rangle \langle [x_{j-1}]_{\theta}|\dots |[x_i]_{\theta}\rangle
\end{equation}

Now make use of the fact that since $T$ commutes with  $H_P$, in all of the transition matrix elements, $\theta$ cannot change. The choice of initial and final states fixes the value of $\theta$ for this computation. Then, use the definition $|[x]_{\theta}\rangle = e^{i \frac{\theta}{2 \pi} \frac{\hat{x}}{l}} |[x]\rangle$ to write the position variables in terms of conventional real numbers and continue with the standard steps associated with computing the path integral (such as inserting complete sets of momentum states and performing Gaussian integrals over momentum). This yields:  

\begin{equation}\label{eq:BlochThetaPI}
\langle [x_f]_{\theta} | U\left(t\right) | [x_i]_{\theta} \rangle = \int_{x\left(0\right) = x_i}^{x\left(t\right) = x_f} Dx\,  \text{exp}^{\left(i \int_{\tau = 0}^{\tau = t} d\tau \left(\frac{1}{2} m  \dot{x}^2  - V\left(x\right) +  \frac{\theta}{2\pi} \frac{\dot{x}}{l}\right)\right)}
\end{equation}

Thus, the Lagrangian now explicitly contains $\theta$. But, this $\theta$ is not a parameter like the mass of the particle - it does not appear in the Hamiltonian. Rather, it reflects the fact that in this theory where there are superselection sectors (or eigenspaces) labelled by the parameter $\theta$, one can compute the time evolution of a specific eigenspace labelled by $\theta$. This is a boundary condition on the time evolution operator and this boundary condition is reflected by a change to the Lagrangian where $\theta$ appears as a parameter. The explicit appearance of $\theta$ in the Lagrangian does not mean that there aren't other eigenspaces of the theory with different values of $\theta$ - the spectrum of  $H_P$ does not change simply because we chose to study the time evolution of a specific sub-space. Thus, if we found a bunch of crystals that started at high energy and all of them ended up at $\theta = 0$, we would still be puzzled by why other values of $\theta$ were not populated.

The description we have provided above is based on the Hamiltonian point of view - we take the Schr\"{o}dinger equation as the fundamental axiom of quantum mechanics and regard the path integral as a solution to this equation. Would we get a different result if we had instead chosen to define the theory entirely using the path integral? For example, we could simply say that the  Lagrangian that we are interested in studying is: 

\begin{equation}
    L = \frac{1}{2} m \dot{x}^2 - V\left(x\right)  +   \frac{\theta}{2\pi}\frac{\dot{x}}{l}
    \label{Eqn:Ltheta}
\end{equation}
In this Lagrangian, $\theta$ certainly looks like a parameter. What is the quantum theory associated with this Lagrangian? Notice that we should not expect anything different from our prior discussion since $V\left(x\right)$ is still a periodic potential. Thus, discrete translations $T$ are still a symmetry of this theory and thus we would expect to see a band structure in the energy eigenstates due to Bloch's theorem. Let us now take \eqref{Eqn:Ltheta} and compute the canonical Hamiltonian from it using the canonical commutation relations between the position $x$ and momentum $p$, $[x, p]=i$. This yields: 

\begin{equation}
    H_{\theta} = \frac{1}{2m}\left(-i \frac{\partial}{\partial x} -\frac{\theta}{2\pi l}\right)^2 + V\left(x\right)
    \label{Eqn:HTheta}
\end{equation}
The $\theta$ in $H_{\theta}$ looks like a parameter. What can we say about the eigenstates of $H_{\theta}$ and $H_P$? It is easy to check that if $\Psi_k\left(x\right)$ is an eigenstate of $H_{\theta}$ then $e^{i\frac{\theta}{2 \pi} \frac{x}{l}} \Psi_k\left(x\right)$ is an eigenstate of $H_P$ with the same eigenvalue. There is thus a 1-to-1 map between these two Hamiltonians and they produce the same dynamics. Specifically, the band structure of $H_P$ is also present in $H_{\theta}$ and any problem associated with picking eigenstates in $H_P$ are also found in $H_{\theta}$. 

This result thus corrects a misconception. It is claimed that pure derivative terms such as $\theta$ in \eqref{Eqn:Ltheta} do not change classical physics but do affect quantum mechanics. This is incorrect - $\theta$ does not affect the overall quantum dynamics either. The theory with a $\theta$ parameter \eqref{Eqn:HTheta} and one without $\theta$ are identical theories with a simple relabeling of states. 

The above path integral in \eqref{eq:BlochThetaPI} (and its analog in QCD) is often thought of as a starting point of the quantum theory as opposed to a solution to the Schr\"{o}dinger equation.  One might be tempted to take this path integral as defining the quantum theory.  However, we see that the restriction of states to a specific $\theta$-sector has not only put a $\theta$ parameter in the action, it has restricted the measure to integrate only over field configurations that are invariant under $x\rightarrow x+ 2\pi l$, which is an unnatural projection in position space.  Once we (or the universe) `measures' the value of $\theta$, then such a path integral is a useful way to parameterize the physical effects of this particular sector. It does not, however, describe the full Hilbert space and cannot be used to ask what might be typical initial conditions for the wave function.

\section{QCD}
\label{sec:QCD}

The discussion around the periodic crystal in Section \ref{sec:pendulum} shares many similarities with the $\theta$ problem in QCD.  To see the analogy, one must construct the Hamiltonian of QCD, including dealing with the gauge invariance of the classical theory.  The classical theory includes a set of gauge fields $A_\mu^a$.  Following Jackiw \cite{Jackiw:1983nv}, the simplest gauge to build a Hamiltonian is temporal or Weyl gauge with $A_0^a = 0$, and one can construct the Hamiltonian out of the field operators $A_i^a$ and their conjugate momenta $\Pi_i^a \equiv - E_i^a$:
\begin{equation}
    \hat{H}_{W} = \int d^3 {\bf x} \, \left( \frac{1}{2}\left(\hat{\bf {E}}^a\cdot\hat{\bf {E}}^a + \hat{\bf {B}}^a\cdot\hat{\bf {B}}^a\right) + \rm{(terms\; with\; quarks)}\right)
    \label{eqn:WeylH-QCD}
\end{equation}
where ${\bf B}^a = \nabla\times {\bf A}^a - (g/2) f^{abc}{\bf A^b \times A^c}$ and $f^{abc}$ are the structure constants of the gauge group.  This Hamiltonian is invariant under 
the remaining spatial gauge transformations.  This remaining gauge group is topologically non-trivial.  This means that there are gauge transformations that cannot be continuously deformed to the identity.  These transformations, called large gauge transformations, can be labeled by an integer $n$.  Thus, there are naive vacuum states, such as that with $\langle 0 | A^{a}_i |0\rangle = 0$, and gauge equivalent vacua $|U_n\rangle\equiv U_n|0\rangle$ generatied by large gauge transformations.  The non-trival topology of the gauge group means that the path in field space from $|U_n\rangle$ to $|0\rangle$ is non-trivial (involves not pure-gauge configurations) and thus has an energy barrier.

The situation is thus similar to that of the case of the crystal. The states $|0\rangle$ and $|U_n\rangle$ mix due to quantum tunneling and the spectrum of QCD develops a band structure.  The energy eigenstates of $H_P$ were labeled by a parameter $\theta$ - similarly, the energy eigenstates of QCD can also labeled by a parameter $\theta$. Since the $|\theta\rangle$ are eigenstates of the full QCD Hamiltonian, no transitions are possible from one value of $\theta$ to a different value, thus each $\theta$  defines a super-selection sector. 

The purpose of this section is not to reproduce Jackiw's lecture notes, but to discuss what natural values of $\theta$ to expect.  So what state should we expect the universe to be in?  A generic initial condition would most naturally be an arbitrary wave functional of fields.  Such a state would be a linear combination of states in different $\theta$ sectors.  A hot early universe would eventually cool down to a linear combination of low-energy states with different $\theta$ values, with matter in each part of the wave functional entangled with the $\theta$ value through interactions (say, with the neutron electric dipole moment).  Similar to the examples of the hydrogen atom with a massive photon or the electron in a crystal, the value of $\theta$ will be chosen probabilistically.  Thus, this cannot be a result of tuning a Hamiltonian parameter to zero, but would be an accident, one not even driven by anthropic reasoning (see, for example, \cite{Banks:2003es,Donoghue:2003vs}).  Similarly, $\theta$ cannot be set to zero by a symmetry argument ({\it i.e.}, imposing CP as a symmetry of the Hamiltonian), in the same way that rotational symmetry does not prevent one from finding a rotationally non-invariant state of the hydrogen atom.

Similar to the construction discussed in Section \ref{sec:pendulum}, if we are solely interested in the physics of a specific $\theta$, one can construct a path integral which time evolves just this sector. In the case of the crystal, we constructed states that were invariant under discrete spatial translations. For QCD, we need states that are invariant under large gauge transformations. These transformations change the winding number of gauge configurations, measured by the operator $G\tilde{G}$.  Using this fact, the Lagrangian to time evolve a specific $\theta$ sector, results in the additional term $\theta G \tilde{G}$ appearing in it. But again, this $\theta$ is not a parameter of the full Hamiltonian.  Rather, it is reflective of the fact that the physics has been restricted to a specific sub-sector of the Hilbert space.

Perhaps one could decide that the path integral that time evolves a specific $\theta$ sector simply defines the full quantum theory, making $\theta$ a fundamental parameter of the time evolution, thus rejecting the Schr\"{o}dinger equation (and thus the Hamiltonian) as the fundamental axiom of quantum mechanics.  However, in the field basis, this is an artificial imposition, similar to that in the crystal example.  It is the full spectrum of the Hilbert space that gives rise to the band structure and thus these $\theta$ vacua.  The path integral is a Green's function that describes the time evolution of a basis, and it is the construction of this basis that makes the $\theta$ vacua meaningful.  One could perhaps try to justify this identification by imposing an identification of states related by large gauge transformations.  This would be equivalent to taking an electron in a crystal and imposing periodic boundary conditions -- {\it i.e.}, putting the electron on a circle.  However, as we show in Appendix \ref{appendix}, the Hilbert space still contains the full phase structure due to the topological non-trivial sectors.

To illustrate these points further, consider the theory of classical electrostatics. This is a linear theory (just like quantum mechanics) and is thus fully described by its Green's functions. One can choose various boundary conditions for the Green's functions and obtain different kinds of solutions to the theory. For example, one can pick free space boundary conditions and obtain the conventional Coulombic Green's function of the theory. This Green's function can be used to obtain the electric potential of any arbitrary charge distribution in space. This general Green's function is analogous to the path integral without $\theta$ in it since it is not restricted to charges in any specific part of  space. Alternately, one can also construct Green's functions by choosing boundary conditions appropriate for the interior of a hollow conductor described by the spatial region $\Omega$.  This Green's function can be used to obtain the electric fields of any charge distribution inside $\Omega$. Further, if we construct Green's functions $G_{1,2}$ for two non-overlapping conducting regions $\Omega_{1,2}$, the charges in $\Omega_1$ cannot influence the fields in $\Omega_{2}$ and vice versa. Thus, we can indeed think of $G_{1,2}$ as ``super-selection'' sectors and these Green's functions are analogous to path integrals with $\theta$ in them since they can only be used to describe fields inside the specific regions $\Omega_{1,2}$. Would we now say that there are an infinite number of distinct theories of electrostatics described by the ``parameters'' $\Omega$, or rather that there is one theory of electrostatics with a wide range of boundary conditions $\Omega$? Since the universe could have produced a wide range of possible conducting shapes, if we happened to live inside a perfectly spherically symmetric box, would we not be confused? Would the imposition of spherical symmetry solve this problem even though the spherically symmetric theory of electrostatics is nevertheless allowed to contain conductors of arbitrary shape? 

Before concluding this section, we clarify certain statements usually made in the context of the Lagrangian regarding $\theta$ and the phase $\phi_q$ of the determinant of the quark mass matrix.  It is well known that all the physical CP violating effects in the QCD vacuum arise from the sum $\bar{\theta} = \theta + \phi_q$.  In a sector with a particular $\theta$, this is clear from the path integral where one can perform a redefinition of the quark fields via a chiral rotation. This rotation shifts the parameters $\theta$ and $\phi_q$ by equal and opposite amounts \cite{Fujikawa:1979ay}.  Thus, physical observables only depend upon this sum. 

If physical observables only depend on $\bar{\theta}$, why are we allowed to talk about $\theta$ separately from $\phi_q$? The Hamiltonian picture provides a clear understanding of the physics. Consider two Hamiltonians. The first, $H$ is the QCD Hamiltonian with a particular phase $\phi_q$.  The second, $H'$ is the same Hamiltonian with the quark field operators redefined by a chiral rotation, and thus has a different quark mass phase, $\phi_q'$.  Both of these describe the theory of QCD coupled to massive quarks. 
Both $H$ and $H'$ are Hamiltonians whose spectrum consists of a band structure with the eigenstates labeled by the parameters $\theta$ and $\theta'$, respectively. Since the physical CP violating effects are set by $\bar{\theta} = \theta + \phi_q$, we can always find $\theta'$ so that $\bar{\theta} = \theta + \phi_q = \theta' + \phi_q'$. That is, there is a map from the spectrum of the Hamiltonian $H$ to the spectrum of $H'$. This is loosely described as shifting the phase from the determinant of the quark mass matrix to $\theta$. But, once we fix the parameter $\phi_q$, the Hamiltonian still contains a spectrum labelled by $\theta$, leading to a set of quantum states each of which will exhibit different CP violation $\bar{\theta} = \theta + \phi_q$. 

Let us also see how the massless quark solution solves the strong CP problem. From the Lagrangian and path integral perspective, there is a redefinition of the quark fields represented by a chiral rotation that shifts the $\theta$ term in the action but leaves the quark mass matrix invariant.  This is because there is a U(1) symmetry in the classical Lagrangian ({\it e.g.}, let's say the up quark is massless -- then this would be proportional to a phase rotation of the right-handed up quark), and thus suggests that the $\theta$ angle is not physical.  However, this solution is in fact a dynamical solution -- the $U(1)$ chiral rotation plays the role of the Peccei-Quinn (anomalous) symmetry.  Since the approximate symmetry is both spontaneously (through chiral symmetry breaking) and explicitly (through the mixed QCD anomaly) broken at the QCD scale, the associated axion is also at that scale -- it's the $\eta'$ \cite{Dvali:2005zk}.  Thus, for all $\theta$ sectors of the theory, the $\eta'$ finds its minimum at zero CP violation.   When the quarks have a non-zero mass, the $\eta'$ receives a mass both from the anomaly and from the explicit non-zero quark masses. Thus, the $\eta'$ has an ``axion quality problem'' wherein its minimum is no longer at the location where $\bar{\theta}$ is canceled.

\section{Gauge Theories and Constraints}
\label{sec:gauge}
The $\theta$ term in the Lagrangian of QCD is not a parameter of the Hamiltonian - it is instead the consequence of restricting time evolution to a specific vacuum sector of the theory. This phenomenon, where the initial conditions of the quantum state being time evolved changes the Lagrangian in the path integral, arises more broadly in gauge theories.  In gauge theories, there are two kinds of classical equations - the constraint equations that impose requirements on the initial states (such as Gauss's law in electromagnetism) and dynamical equations. But, the classical equations are not the fundamental equations of nature - rather, they are identities that follow from quantum mechanics. Quantum evolution is described by a single first order differential equation - the Schr\"{o}dinger equation. This equation can time evolve any initial state, including states that violate the classical constraints. The information about states that violate constraints does not live in the Hamiltonian. But, since the Lagrangian emerges as a solution to the dynamical equations, it can be impacted by boundary conditions. As we will see below, the time evolution of quantum states that violate the constraint equations can appear in the Lagrangian in the form of Lorentz breaking background fields. But, these background fields are not new degrees of freedom - rather they are indicative of the fact that we are choosing to time evolve a specific kind of initial state - this is exactly like the case of $\theta$. For simplicity, we will restrict our arguments to electromagnetism, but they can be broadly extended to both general relativity and QCD. 

Electromagnetism is the theory of an abelian gauge boson $A_{\mu}$. We will work in the Schr\"{o}dinger picture. One needs a Hamiltonian constructed from the gauge field and its conjugate momentum. Further, due to gauge freedom, a gauge needs to be picked to define the Hamiltonian. Pick the Weyl gauge where $A_0 = 0$. The physical degrees of freedom are thus the spatial components $A_i$ of the gauge boson and the conjugate momentum $\hat\Pi_i$. Using the fact that $\hat\Pi_j = -\hat E_j$, we have
\begin{equation}
[\hat{A_j}\left(\bf{x}\right), \hat{E_{j'}}\left(\bf{x'}\right)] = -i\, \delta\left({\bf x - x'}\right) \delta_{jj'}
\end{equation}
The Hamiltonian constructed from these operators is: 
\begin{equation}
    \hat{H}_{W} = \int d^3 {\bf x} \, \left( \frac{1}{2}\left(\hat{\bf {E}}\cdot\hat{\bf {E}} + \hat{\bf {B}}\cdot\hat{\bf {B}}\right) + \hat{\bf {J}}\cdot\hat{\bf {A}}   + \hat{\cal H}_{J}\right)
    \label{eqn:WeylH}
\end{equation}
where $J$ and $\hat{\cal H}_J$ denote the current and the Hamiltonian of the current. 

Much like the operator $T$ that commuted with $H_P$ in Section \ref{sec:pendulum}, the operator $\hat{G} = \nabla\cdot\hat{\bf{E}} - \hat{J}^0$ commutes with $\hat{H}_W$. It is thus possible to describe the time evolution generated by $\hat{H}_W$ in terms of the eigenstates of $\hat{G}$. These states have the form: 
\begin{equation}
\hat{G}\left({\bf x}\right)|\Psi_{s}\rangle = J^{0}_s\left({\bf x} \right) |\Psi_s\rangle
\label{Eqn:Gauss}
\end{equation}

Notice that $\hat{G}\left({\bf x}\right)$ is the Gauss law operator. Eigenstates of $\hat{G}$ that have zero eigenvalue are quantum states that preserve Gauss's law. However, the operator $\hat{G}$ also allows for eigenstates which have a non-zero eigenvalue. In these states, Gauss's law does not hold - the divergence of the electric field need not exactly match the charge density of the current. This difference is captured by the eigenvalue $J^{0}_s\left({\bf x}\right)$ - this is simply a c-number function. It is {\bf not} a new degree of freedom or a new parameter - it is simply an eigenvalue of the operator $\hat{G}$, analogous to $\theta$ in Section \ref{sec:pendulum}. For the pendulum or for QCD, we do not restrict the Hilbert space to the sector of the theory where $\theta = 0$. There is a tendency to do so for electromagnetism (and other gauge theories) due to the desire to preserve Gauss's law. But, from the quantum perspective, this restriction is unnecessary. Is there anything logically wrong in using the Schr\"{o}dinger equation to evolve a state with $J^{0}_s \left({\bf x}\right) \neq 0$?  

First, note that the Hamiltonian $\hat{H}_{W}$ is invariant under the transformation $\vec{\hat{A}}\rightarrow \vec{\hat{A}} + \nabla \alpha\left({\bf x}\right)$ for any c-number function $\alpha\left({\bf x}\right)$ of space - this is an operator statement, independent of any choice of quantum state that is being time evolved by $\hat{H}_{W}$. This symmetry forbids the generation of operators of the form $\vec{A}.\vec{A}$ that would indicate a mass for the photon. Thus, in the absence of any other currents being turned on, this theory describes a massless photon even when $J^{0}_s \neq 0$. 

At this stage, one can construct a path integral that solves the time evolution generated by $\hat{H}_{W}$. The canonical procedure will produce the Lagrangian: 
\begin{equation}
\mathcal{L}_{W} = -\frac{1}{4} F_{\mu\nu}F^{\mu\nu} + \mathcal{L}_J - {\bf A}\cdot {\bf J}
\end{equation}
where $\mathcal{L}_{J}$ is the Lagrangian for the current $J$. Notice that this Lagrangian is not Lorentz invariant - this is unsurprising since we picked  time slices to define the Hamiltonian and picked a gauge $A_0 = 0$ with respect to that time slicing. This breaking of Lorentz is however innocuous - the theory is causal, the photon is massless and the speed of light is still the same for all massless particles. These are the physical consequences of Lorentz invariance that we actually care about. 

Notice that $J^{0}_s\left({\bf x}\right)$ does not enter this Lagrangian. This is also unsurprising - the path integral was constructed to solve the time evolution of the entire Hilbert space of the theory and it was not restricted to any specific sub-space. Indeed, if one obtains the classical equations of motion from this Lagrangian (or, more formally, by using the Schwinger-Dyson procedure), one does not obtain Gauss's law since there are no variations associated with $A_0$ which has been fixed to be 0. Therefore, the fact that one can evolve any quantum state (including states that violate Gauss's law) using $\hat{H}_{W}$ is reflected in the fact that the classical equations that follow as an identity from this quantum dynamics lack the constraint equation that would have forbidden the possibility of evolving states that violate the Gauss's law constraint. This is similar to eq.~\eqref{Eqn:NoThetaL} where $\theta$ did not appear in the Lagrangian of the crystal when we were interested in constructing the time evolution of any state in the Hilbert space with arbitrary values of $\theta$. 

Let us now restrict our attention to a specific sector of electromagnetism with a specific eigenvalue $J^{0}_s\left({\bf x}\right)$. That is, we consider states $|[A]_s\rangle$ that are eigenvalues of $\hat{G}$ and compute the transition matrix elements: 

\begin{equation}
    \langle [A_f]_s| e^{-i \hat{H}_{W}t} | [A_i]_s\rangle = \langle [A_f]_s| e^{-i \hat{H}_{W} \delta t}\dots e^{-i \hat{H}_{W} \delta t} | [A_i]_s\rangle
\end{equation}

To proceed further, we need to insert a resolution of unity in between the infinitesimal time steps $\delta t$ in the above approximation. But, since $\hat{G}$ commutes with $\hat{H}_W$, we know that there are no transitions possible between states of different eigenvalues $J^{0}_s$. It is thus sufficient to insert the states: 

\begin{equation}
|[A]_s\rangle = \hat{P}_s |A\rangle
\end{equation}
where $\hat{P}_s$ is the projector from the field basis state $|A\rangle$ to the appropriate eigenspace. This projection operator is implemented as a delta function: 
\begin{equation}
\hat{P} = \Pi_{t,{\bf x}} \,  \delta \left(\hat{G}\left({\bf x}\right) - J^{0}_{s} \left( {\bf x}\right) \right) 
\label{eqn:delta}
\end{equation}

We insert these states at each point in the path integral and implement the delta function using its integral representation

\begin{equation}
\delta \left({\bf \nabla}\cdot{\bf E}\left( {\bf x}, t\right) - {J}^0 \left( {\bf x}, t\right) - J^{0}_{s}\left({\bf x}\right) \right)  = \int DA_{0} \,  e^{i \delta t \int d^3 {\bf x} \, A_0 \left( {\bf x}, t\right) \left({\bf \nabla}\cdot {\bf E}\left({\bf x}, t\right) - J^{0}\left( {\bf x}, t\right) -J^{0}_{s}\left({\bf x} \right)\right) }
\end{equation}

With these insertions and proceeding using standard techniques \cite{Kaplan:2023fbl}, we  get the Lagrangian

\begin{equation}
\mathcal{\tilde{L}_{EM}} = -\frac{1}{4}F_{\mu \nu}F^{\mu \nu} +  A_{\mu} J^{\mu} + A_{\mu}J_{s}^{\mu}+ \mathcal{L}_{J}
\label{eqn:ourQED}
\end{equation}
where the field $J_{s}^{\mu}$ is the background classical field $J_{s}^{\mu}=\left(J_{s}^0\left( {\bf x}\right), 0, 0, 0\right)$. In this Lagrangian, $J_s$ looks like a parameter - a background classical field that was added to the Lagrangian in a seemingly ad-hoc way, similar to how $\theta$ would naively appear as a  parameter in the Lagrangian of the crystal or QCD. But, the origin of $J_s$ in this case is identical to the origin of $\theta$ terms. Neither of them exist as parameters in the Hamiltonian. They do not appear in the Lagrangian that evolves generic states of the theory. Their origin does not represent new physics or new degrees of freedom - they are merely states that are part of the full Hilbert space of the theory. Since their origin is coupled to the existence of operators that commute with the full Hamiltonian, their eigenspaces are super-selection sectors. Thus, one can restrict time evolution to any specific super-selection sector and this restriction results in these terms appearing as parameters in the Lagrangian. 

Finally, just like in the case of the crystal and QCD, we can map the physics of the above Hamiltonian to that of a Hamiltonian where the background appears. States with background charge $J^0_s({\bf x})$ are constructed as
\begin{equation}
    |[A]_s\rangle = e^{i\int d^3{\bf x}\,  {\bf A}\cdot\nabla \Phi_s \, }| 0 \rangle  \equiv U_s |0\rangle\,,
\end{equation}
where $\nabla^2\Phi_s= J^0_s({\bf x})$, and where $| 0 \rangle$ is the usual vacuum of electromagnetism, i.e. with an absence of background charge. This is a unitary transformation that effectively shifts background charge between states,
\begin{equation}
      U_{s'}|[A]_s\rangle= |[A]_{s+s'}\rangle 
\end{equation}
where $|[A]_{s+s'}\rangle $ has background charge $J^0_s({\bf x})+J^0_{s'}({\bf x})$. Under this unitary transformation the Hamiltonian transforms as
\begin{equation}
   \hat H_W \to U^\dagger_s \hat H_W U_s = \int d^3{\bf x} \left(\frac{1}{2} \left((\hat {\bf E} - \nabla \Phi_s)^2 + \hat {\bf B}^2\right) + \hat{\bf {J}}\cdot\hat{\bf {A}}   + \hat{\cal H}_{J}\right)
\end{equation}
In parallel with the crystal and QCD cases, while $\nabla \Phi_s$ appears in this Hamiltonian as a parameter, the dynamics of this theory is the same the dynamics of $\hat{H}_W$ when one includes the full Hilbert space of quantum states, including those that violate Gauss's law. There is also an additional similarity between states that violate Gauss's law and the $\theta$ term in 1+1 dimensional QED. In this theory, the $\theta$ term is simply a source-less background electric field that arises from the choice of vacuum of the electromagnetic vector potential $A$. States that violate Gauss's law are also source-less background electric fields that reflect a different choice of initial condition for $A$.

\section{Conclusions}
\label{sec:conclusions}

In classical mechanics, the Hamiltonian and the Lagrangian are  functions that are Legendre transformations of each other. They are treated on equal footing and parameters in one can be mapped to parameters in the other. This is not the case in quantum mechanics. In quantum mechanics, the Hamiltonian is an operator which defines the dynamical equation of motion whereas the Lagrangian is still a  function that appears in the path integral. The path integral  is a solution to the equation of motion. Solutions to differential equations require initial conditions and thus, it is possible, in principle, for the Lagrangian to depend explicitly on the initial conditions. In linear quantum mechanics, no such dependence is possible in the Hamiltonian. In this paper, we have displayed explicit examples of this phenomenon in the case of the $\theta$ term of QCD and states that violate Gauss's law in electromagnetism. 

Given this understanding, what can we say about solutions to the strong CP problem? The $\theta$ vacua exist ({\it i.e.} cannot be eliminated by imposing parity as a symmetry on the Hamiltonian) and the value of $\theta$ chosen by our universe is the probabilistic outcome of measurement. This randomly chosen value needs to be nearly exactly canceled by an actual parameter in the theory - namely, the argument of the determinant of the quark mass matrix. A fine tuned cancellation would require tuning an initial parameter of the theory against the probabilistic outcome of measurement. 


Naively, there might seem to be at least two strategies that are theoretically possible.  The first is the dynamical mechanism of the axion. In this case, no matter what the initial value of $\bar{\theta}$, the axion field dynamically adjusts itself to cancel $\bar{\theta}$. Our result considerably strengthens the theoretical case for an axion. Moreover, given that that the initial value of $\theta$ is a random value produced as a result of measurement, it is exceedingly likely that this random value will be different from the initial value of the axion field. Thus, if the axion exists, it is likely to have a cosmological abundance, motivating the search for axion dark matter. Note that we include the possibility of a massless quark into this class of solution. This is because, as we have seen,  when there is a massless quark, the $\eta'$ effectively acts as an axion and dynamically cancels $\theta$.  Indeed, this is the same phenomenology that occurs in composite axion models where one considers a massless quark that is confined under some other gauge group. If the confinement scale is high, this massless quark would be consistent with observations, but the confined sector results in the existence of a composite axion \cite{Kim:1984pt} at low energies. 

The second possibility is the possibility of significantly altering the spectrum of QCD so that either the $\theta$ vacua are not eigenstates of the Hamiltonian or their spectrum is dramatically altered. Proposals along the latter vein include dynamics that leads to an accumulation of vacua around $\theta = 0$ \cite{Dvali:2005zk} which may result in the high energy QCD states preferentially decaying to these states with small values of $\theta$. Alternatively, one may consider the possibility of a non-abelian gauge theory that is not invariant under large gauge transformations (thus eliminating the $\theta$ vacua) but is still invariant under a smaller set of gauge transformations that may be sufficient to maintain the masslessness of the gauge field.  In both of these cases, while $\theta$ would be small, there would still be a CP violating phase from the determinant of the quark mass matrix. But, this could be eliminated by imposing CP as an additional symmetry of the theory. 

In the case of electromagnetism (and more broadly, gauge theories), we see that we can start with a Hamiltonian that allows for the existence of a Lorenz invariant vacuum (such as $\hat{H}_W$). Nevertheless, this Hamiltonian contains states that break Lorentz symmetry. The space of states that break Lorentz symmetry includes states that violate constraints such as Gauss's law. Generically, just as in the case of $\theta$ vacua, we would expect to live in a universe where the quantum state violates Gauss's law. Why do we then live in a universe where Gauss's law appears to be well preserved? Similar to the case of the strong CP problem, this issue has to be addressed dynamically. Unlike the case of the strong CP problem where the axion has not been experimentally discovered, the dynamical mechanism that can suppress violations of Gauss's law is known and empirically verified - namely, the expansion of the universe.


\section*{Acknowledgements}
We thank Asimina Arvanitaki, Savas Dimopoulous, Peter Graham and Hitoshi Murayama  for fruitful discussions. This work was supported by the U.S.~Department of Energy~(DOE), Office of Science, National Quantum Information Science Research Centers, Superconducting Quantum Materials and Systems Center~(SQMS) under Contract No.~DE-AC02-07CH11359. D.E.K.\ and S.R.\ are supported in part by the U.S.~National Science Foundation~(NSF) under Grant No.~PHY-1818899.
S.R.\ is also supported by the~DOE under a QuantISED grant for MAGIS.
The work of S.R.\  was also supported by the Simons Investigator Award No.~827042. T.M.\ is supported by the World Premier International Research Center Initiative (WPI) MEXT, Japan, and by JSPS KAKENHI grant JP22K18712. 

\begin{appendices}
\section{Pendulum in Gravity}
\label{appendix}

\begin{figure*}[t!]
\centering 

\includegraphics[width=.5\linewidth]{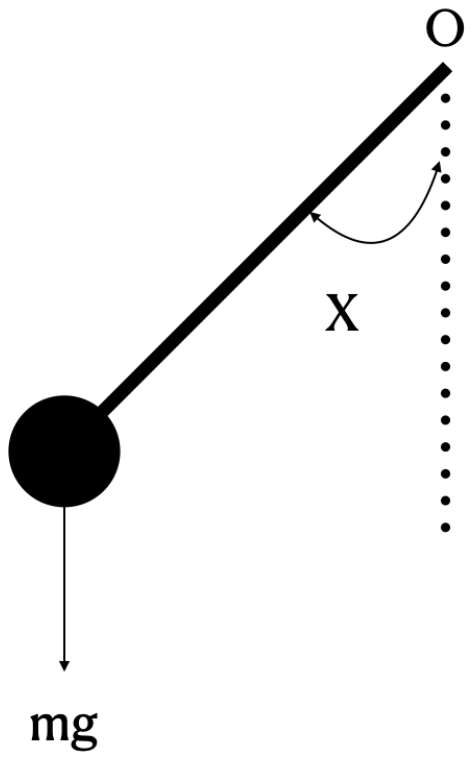}

\caption{A bob is attached to a rigid rod. The angle between the bob and the vertical is x. The Hilbert space of this pendulum is the space of all continuous quantum states that can be defined on the circle. Since quantum states that differ by an overall phase are in fact the same quantum state even if they are distinct complex functions, the Hilbert space of this theory is larger than being the set of all continuous complex functions on the circle. } \label{Fig:Pendulum}
\end{figure*}

We briefly discuss the quantum mechanics of a pendulum in gravity that is often discussed in the context of $\theta$ vacua. Consider a particle of mass $m$ attached to a rigid rod oscillating in a gravitational field as shown in figure \ref{Fig:Pendulum}. Denote the angle between the pendulum and the vertical by $x$. The rigidity of the rod allows it to swing all the way from $x = 0$ to $x = 2 \pi$. The potential experienced by this particle is $V\left(x\right) = - mg \cos\left(x\right)$, similar to the periodic potential of a particle in a crystal as discussed in Section \ref{sec:pendulum}. We expect the physics of these two systems to be similar. Let us see how this works out. 

Begin by constructing the Hilbert space of the theory. The pendulum moves along a circle. One cannot cover the circle by a single global co-ordinate patch -- we can pick two patches, one that covers the top part of the circle and the other the bottom half. The Schr\"{o}dinger equation can be written using these coordinate patches and quantum states that obey the respective equations can be obtained. We need to specify boundary conditions at the patches where the states overlap. We require the quantum states to be continuous at these boundaries.  But, note that since two quantum states that differ by an overall phase are in fact the same quantum state, these quantum states are not elements of the space of complex functions, but are rather elements of the space of complex projective functions. Functions that are continuous in the projective space, and are thus allowed quantum states, can be discontinuous in the  space of complex functions. The Hilbert space of this theory is thus larger than simply the space of periodic complex functions on the circle.

Let us see how to construct this Hilbert space. We wish to represent the quantum states on the circle using the position space basis states $|[x]\rangle$ defined between $0 \leq [x] \leq 2 \pi$ with the end points identified. But,  the state $|[x]\rangle$ and $e^{i \alpha}|[x]\rangle$ are the same physical state even though they are represented by two different complex functions. The states $|[x]\rangle$ and $e^{i \alpha}|[x]\rangle$ are members of the same equivalence class $|[[x]]\rangle$, $|[x]\rangle \equiv e^{i \alpha}|[x]\rangle$. To construct the Hilbert space, we can pick any element of each of these equivalence classes $|[[x]]\rangle$. The operators we use to define  the theory should be insensitive to the choice of elements in the equivalence class. The position basis is thus the set $e^{i\alpha\left([x]\right)} |[x]\rangle$ where we have picked an element from every equivalence class $|[[x]]\rangle$. The phase $\alpha\left([x]\right)$ can vary with position. Focus on the momentum operator $-i \frac{d}{dx}$. Clearly, this operator acts differently on the complex function $\Psi\left(x\right)$ and $e^{i \alpha\left(x\right)} \Psi\left(x\right)$, even though the latter is obtained by picking a different set of representatives from the equivalence classes of position vectors. To define the quantum theory properly, we should introduce a non-dynamical gauge field into the theory to absorb the choice of gauge in the Hilbert space. The momentum operator is thus: 

\begin{equation}
-i \frac{\partial}{\partial x} \rightarrow -i \frac{\partial}{\partial x} + A\left(x\right)
\end{equation}
where the gauge field $A\left(x\right)$ shifts as $A\rightarrow A - \partial \alpha$ when we choose a different set of representatives so that $\Psi\left(x\right) \rightarrow e^{i \alpha\left(x\right)} \Psi$. This is simply a covariant derivative on the Hilbert space, that suitably changes when different phases are picked for the basis. When we do quantum mechanics on $R^1$ (or any topologically trivial manifold), we do not worry about this since any choice of $\alpha$ can be gauged away.

But, the circle is topologically non-trivial, leading to a more complex structure on the Hilbert space. Let  $|0\rangle$ denote the set of basis states  $|[x]\rangle$ where all the states have exactly the same phase. Alternately, we can choose a different set of basis states denoted by $|n\rangle$ where the basis states are of the form $e^{i \alpha_{n}\left([x]\right)}|[x]\rangle$ with the requirement that the function $\alpha_{n}\left([x]\right)$ is continuous in the interval $0 \leq [x] \leq 2 \pi$ with $\alpha_{n}\left([2 \pi]\right) - \alpha_{n}\left([0]\right) = 2 \pi n$ for integer $n$. On the circle, basis choices $\alpha_{n}$ and $\alpha_{m}$ with $n \neq m$ are distinct - they cannot be smoothly deformed into each other. The circle thus has topologically distinct sets of basis vectors $|n\rangle$, with $n=0,\pm1,\pm2\ldots$. That is, we have position vectors $|[x]_{m}\rangle = e^{i \alpha_{m}\left([x]\right)}|[x]\rangle$ and  $|[x]_{n}\rangle = e^{i \alpha_{n}\left([x]\right)}|[x]\rangle$ which are distinct when $m \neq n$, significantly enhancing the size of the  Hilbert space. The basis states $|0\rangle \dots |n\rangle$ are analogous to the vacua that were naively constructed for QCD where the gauge field states $|0\rangle, \dots |U_n\rangle$ are topologically distinct choices to construct the Hilbert space of the theory. But, exactly like the case in QCD, quantum tunneling mixes all the basis states $|0\rangle \dots |n\rangle$ together. 

How do we construct the energy eigenstates of this system given this enhanced Hilbert space? Cover the circle with two co-ordinate patches and  match the quantum state smoothly where the patches overlap. Take the first co-ordinate patch $P_1$ to be all the points on the circle except the point $x = 0 \equiv 2 \pi n$ and the second to be all the points except for $x =  \pi \equiv \left(2 n + 1 \right) \pi$ for integer $n$.  The match needs to be continuous in the space of complex projective functions and not the space of complex functions. The $\theta$ vacua arise  while performing this match.  Now, each co-ordinate patch is just a line segment. In each of these co-ordinate patches, we can describe the Hilbert space using the position basis - specifically, we can pick representatives $e^{i \alpha\left(x\right)}|x\rangle$ from the equivalence class $|[[x]]\rangle$. On each patch, we can smoothly deform any gauge choice  to a constant. Without loss in generality, take the constant in $P_1$ to be $0$ and the constant in $P_2$ to be $\theta$. The Hamiltonian then takes the form: 

\begin{equation}
H_1 = \frac{1}{2m} \left( -i \frac{\partial }{\partial x}\right)^2 + V\left(x\right)
\end{equation}

\begin{equation}
H_2 = \frac{1}{2m} \left( -i\frac{\partial}{\partial x} - \frac{\theta}{2 \pi}\right)^2 + V\left(x\right)
\end{equation}
where $H_{1,2}$ are the Hamiltonians in the first and second patch respectively. As described earlier, the conjugate momentum of $x$ is shifted by $\theta$ to account for the gauge freedom in the choice of basis elements. Consider eigenfunctions $\Psi_{1,2}$ of $H_{1,2}$ with the same energy $E$. These are related by $\Psi_{1}\left(x\right) = e^{i \frac{\theta x}{2 \pi} } \Psi_{2}\left(x\right)$ in the regions where $\Psi_{1,2}$ are both defined. Now $\Psi_{2}$ is defined at the point $x = 0$ while $\Psi_1$ is not. But, using the relation  $\Psi_{1}\left(x\right) = e^{i \frac{\theta x}{2 \pi} } \Psi_{2}\left(x\right)$, we can compute $\text{lim}_{x\rightarrow 0} \Psi_1\left(x\right) = \text{lim}_{x\rightarrow 0} e^{i \frac{\theta x}{2 \pi}} \Psi_2\left(x\right)$ and  $\text{lim}_{x\rightarrow 2 \pi} \Psi_1\left(x\right) = \text{lim}_{x\rightarrow 2\pi} e^{i \frac{\theta x}{2 \pi}} \Psi_2\left(x\right)$. Since $\Psi_{2}$ is a well defined function at $x = 0 \equiv 2 n \pi$, we have $\text{lim}_{x\rightarrow 2 \pi} \Psi_1\left(x\right) = e^{i \theta} \text{lim}_{x\rightarrow 0} \Psi_1\left(x\right)$. We thus see that the energy eigenstates of this system are states that obey the equation $H_1 \Psi_1 = E \Psi_1$ with the boundary condition $\text{lim}_{x\rightarrow 2 \pi} \Psi_1\left(x\right) = e^{i \theta} \text{lim}_{x\rightarrow 0} \Psi_1\left(x\right)$. These are precisely the energy eigenstates of the periodic crystal. $\theta$ looks like a boundary condition or parameter choice - but it isn't. $\theta$ labels the various possible choices of position state basis vectors that could have been picked in $P_2$ and it is thus once again tied to picking the right elements in the Hilbert space.

One can ask if such $\theta$ vacua exist if the circle is obtained from $R^3$, as is the case in our universe, as $R^3$ is simply connected and the Hilbert space can be constructed from the position basis in the conventional manner. 
A simple physical example is a charged particle confined to a loop of wire.  Naively, the particle could have wave functions that are exactly periodic (and thus trivial under the translation operator $T$).  One could then create the same setup with a magnetic flux through the loop of wire (but zero magnetic field in the region of the wire).  Famously, there is an Aharonov-Bohm phase when the particle wave function is translated around the wire.  Thus, decoupling all else, this would be a quantum theory of a particle on a circle with a particular phase boundary condition ({\it i.e.}, confined to a particular $\theta$ sector of the Hilbert space).  Finally, one could create the same loop in a quantum state with a linear combination of magnetic fluxes.  States of the particle on the wire would now live in a linear combination of theta sectors, which are ultimately connected by a decoupled part of the theory -- the source of the magnetic field.

\end{appendices}

\bibliographystyle{unsrt}
\bibliography{sample}

\end{document}